\documentclass[prl,aps,twocolumn,longbibliography]{revtex4-2}

\usepackage[paperwidth=210mm,paperheight=297mm,centering,top=2.1cm,bottom=2.5cm,right=1.9cm,left=1.8cm]{geometry}

\usepackage{titlesec}
\titlespacing*{\subsection} {0pt}{3ex}{3ex}

\pdfoutput=1

\newcommand{\trf}{t_{\textrm{rf}}}
\newcommand{\kB}{k_{\textrm{B}}}

\newcommand{\knb}{k_n}
\newcommand{\Enb}{E_n}
\newcommand{\UD}{U_{\textrm{D}}}

\newcommand{\Bres}{B_{\textrm{res}}}
\newcommand{\Ep}{E_{\textrm{p}}}

\newcommand{\ab}{a_{\textrm{b}}}

\newcommand{\Tc}{T_{\textrm{c}}}

\usepackage{float}
\usepackage{fourier}
\usepackage{amsmath,amssymb}
\usepackage{mathtools}
\usepackage{amsthm}
\usepackage{empheq}
\usepackage{cases}
\usepackage{times}
\usepackage{upgreek}
\usepackage{psfrag} 
\usepackage{latexsym} 
\usepackage{amstext}
\usepackage{amsfonts}
\usepackage{amsxtra} 
\usepackage{dcolumn}
\usepackage{dcolumn}
\usepackage{textcomp}
\usepackage{graphicx}
\usepackage{bm}
\usepackage{braket}
\usepackage{units}
\usepackage[svgnames,dvipsnames]{xcolor}

\definecolor{myColor}{rgb}{0.02,0.12,0.3}
\definecolor{myColor2}{rgb}{0.1,0.2,0.65}
\definecolor{myciteColor}{rgb}{0.39,0.7,0.89}
\usepackage[colorlinks=true,citecolor=myColor,linkcolor=myColor,urlcolor=myColor]{hyperref}

\def\be{\begin{equation}}
\def\ee{\end{equation}}

\def\nobreakbefore{%
  \relax\ifvmode\else
    \ifhmode
      \ifdim\lastskip > 0pt\relax
        \unskip\nobreakspace
      \else % added to put a ~if no space was typed.
        \nobreakspace
      \fi
    \fi
  \fi
}
\let\oldcite\cite
\renewcommand\cite{\nobreakbefore\oldcite}

\makeatletter
\def\@fnsymbol#1{\ensuremath{\ifcase#1\or *\or \dagger\or \ddagger\or
   \mathsection\or \mathparagraph\or \|\or **\or \dagger\dagger
   \or \ddagger\ddagger \else\@ctrerr\fi}}
\makeatother

\begin{document} 
 
\title{
Fate of an impurity strongly interacting with a thermal Bose gas
}
\author{Ji\v r\' i~Etrych$^{1}$, Sebastian~J.~Morris$^{1}$, Simon M.~Fischer$^{1}$, Gevorg~Martirosyan$^{1}$, Christopher J.~Ho$^{1}$, Moritz~Drescher$^{2}$, Manfred~Salmhofer$^{2}$,
Zoran~Hadzibabic$^{1}$, Tilman~Enss$^{2}$, and Christoph~Eigen$^{1,\ast}$}
\affiliation{
$^1$\,Cavendish Laboratory, University of Cambridge, J. J. Thomson Avenue, Cambridge CB3 0HE, United Kingdom\\
$^2$\,Institut für Theoretische Physik, Universität Heidelberg, 69120 Heidelberg, Germany}

\begin{abstract}
We spectroscopically study mobile impurities immersed in a homogeneous bosonic bath (a box-trapped Bose gas), varying the bath temperature and the strength of impurity-bath interactions.
We compare our results to those for a quasipure Bose--Einstein condensate (BEC), and find that for strong impurity-bath interactions, the spectra narrow with increasing temperature, while the impurity energy shift is suppressed. Near the critical temperature for condensation, many-body effects still play an important role, and only for a nondegenerate bath the system approaches the classical Boltzmann-gas behavior. The key spectral features are reproduced within the theory of the ideal Bose polaron.
\end{abstract}
\maketitle
\vspace{-1em}

An impurity immersed in a quantum bath is a fundamental setting in many-body physics that, in spite of its apparent simplicity, features complex emergent behavior. A key simplification often lies in treating the impurities dressed by the excitations of the bath as polaron quasiparticles. This notion originates from descriptions of electrons moving through a polarizable crystal~\cite{Landau:1933,Pekar:1946a}, but nowadays extends to numerous scenarios in condensed matter physics and beyond~\cite{Alexandrov:2008,Franchini:2021}.

In cold-atom experiments, impurity physics can be studied in highly imbalanced two-component mixtures, where the minority component acts as impurities and the majority component as the bath~\cite{Scazza:2022,Baroni:2024b,Grusdt:2025,Massignan:2026}. 
A key advantage of these experiments is the high degree of tunability: the bath density $n$ and temperature $T$ are readily controllable, and the strength and sign of the impurity-bath interactions, characterized by the s-wave scattering length $a$, can be tuned using Feshbach resonances.
In the high-temperature regime, the mixture can be described as a classical Boltzmann gas, and the thermal de Broglie wavelength $\lambda\propto 1/\sqrt{T}$ sets the relevant interaction parameter $a/\lambda$.
Instead, in the low-temperature regime, the impurity behavior depends on the quantum statistics of the bath, and is generally governed by the interaction parameter $\knb a$, where $\knb=(6\pi^2 n)^{1/3}$\cite{Scazza:2022,Baroni:2024b,Grusdt:2025,Massignan:2026}.

For a fermionic bath, long-lived attractive and repulsive polaron quasiparticles  exist for temperatures well below the Fermi temperature~\cite{Schirotzek:2009,Kohstall:2012}, but the quasiparticle description breaks down as $T$ is increased~\cite{Yan:2019}.
The situation is more intricate for a bosonic bath: in the limit of vanishing intrabath interactions ($k_n\ab \to 0$, where $\ab$ is the intrabath scattering length), there is no overlap between the ground state of the system with and without the impurity; this is known as the bosonic orthogonality catastrophe~\cite{Yoshida:2018,Guenther:2021}. 
When the bath is weakly repulsive and the reduced temperature $T/\Tc\ll1$, where $\Tc$ is the critical temperature for Bose--Einstein condensation, well-defined attractive and repulsive polarons exist at sufficiently low $|a|$\cite{Jorgensen:2016,Hu:2016,Skou:2021,Skou:2022,Cayla:2023,Etrych:2025},
but in the strongly interacting regime the quasiparticle picture breaks down and the spectra are intrinsically broad~\cite{Etrych:2025}. 
The fate of the system at nonzero $T$, including the effect of the BEC phase transition, is subject to ongoing theoretical debate~\cite{Schmidt:2016,Guenther:2018,Pastukhov:2018,Dzsotjan:2020,Field:2020,Pascual:2021,Drescher:2024}, with previous experiments limited to strong attractive interactions~\cite{Yan:2020}. 
Furthermore, in contrast to fermions, strongly interacting bosons are unstable due to three-body loss (due to the absence of Pauli blocking), making it challenging to investigate their equilibrium properties.

\begin{figure*}[t!]
\centerline{\includegraphics[width=1.0\textwidth]{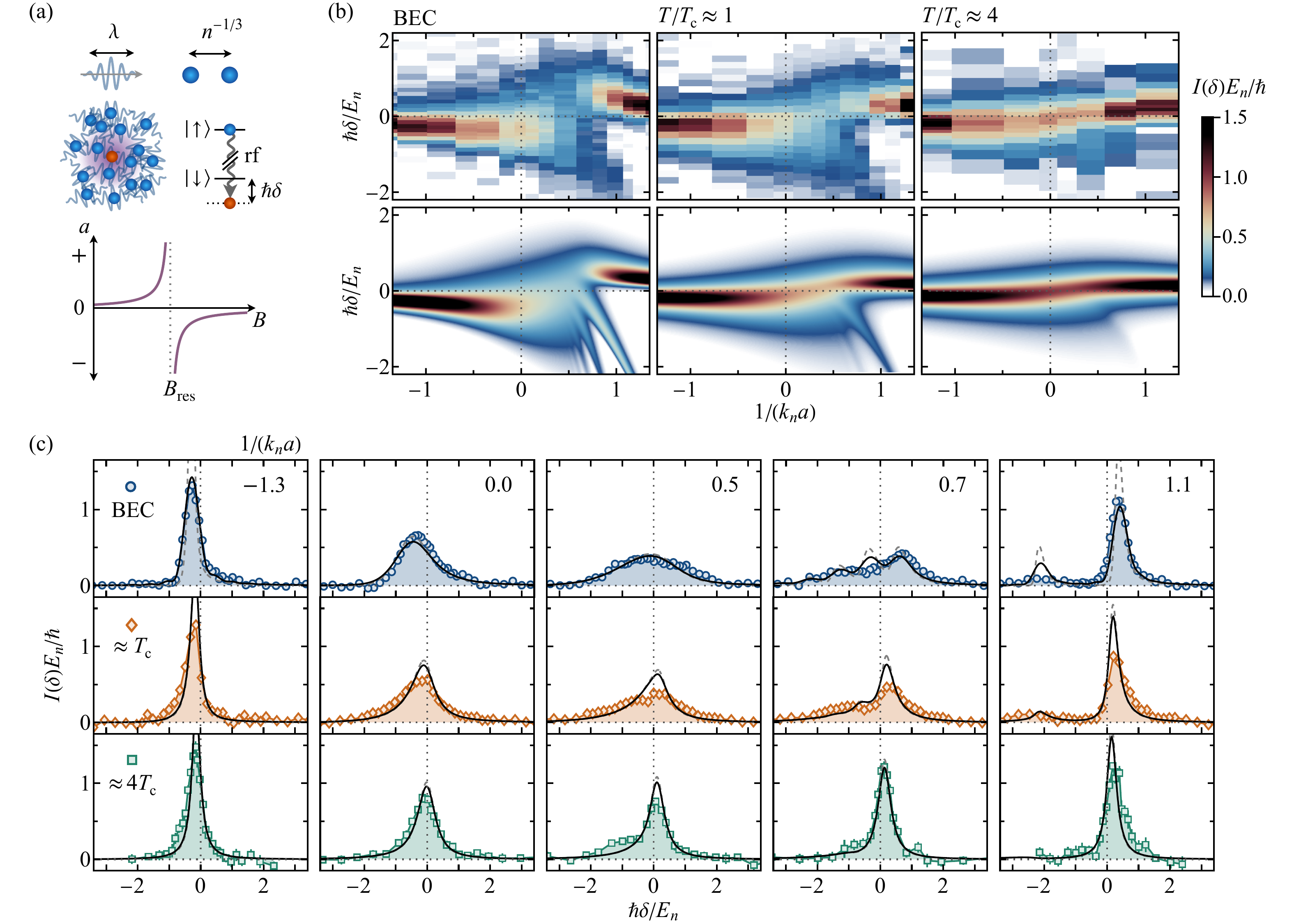}}
\caption{
Injection spectroscopy of impurities in a bosonic bath.
(a)~We start with a spin-polarized Bose gas, with tuneable thermal wavelength $\lambda\propto 1/\sqrt{T}$ and interparticle separation $n^{-1/3}\propto1/\knb$, set by the gas temperature $T$ and density $n$ (top).
We use rf pulses of variable detuning $\delta$ to inject a small fraction of impurities, mapping out the spectral response $I(\delta)$ (middle).
The impurity-bath interactions, characterized by the interstate scattering length $a$, are tuneable by changing the magnetic field $B$ in the vicinity of a Feshbach resonance (bottom).
(b)~Experimental impurity spectra (top) and ideal-polaron calculations (bottom) in the strong-coupling regime for three different baths: an initially quasipure Bose--Einstein condensate (BEC) with $n\approx 12\upmu$m$^{-3}$~\cite{Etrych:2025}~(left), a Bose gas near the critical temperature for condensation $\Tc$ [$T\approx130$\,nK, $n\approx5\,\upmu$m$^{-3}$]~(middle), and a thermal gas at $T/\Tc\approx4$~ [$T\approx260$\,nK, $n\approx1.7\, \upmu\,$m$^{-3}$]~(right).
(c) Example experimental (symbols) and theoretical (solid lines) injection spectra for our three bath temperatures at different $1/(\knb a)$ (left to right), corresponding to vertical cuts through the panels in Fig.~\ref{fig1}(b). In the strong-coupling regime, the spectral response narrows with increasing $T$. 
Note that we have convolved the theoretical spectra with the residual Fourier broadening (arising from the finite rf-pulse duration); for comparison, the dashed lines (in most cases indistinguishable from the solid lines) show the unconvolved spectra.
}
\label{fig1}
\end{figure*}

In this Letter, we study impurities injected into a box-trapped Bose gas, mapping out the impurity spectrum across the strong-coupling regime for different $T/\Tc$.
We compare our measurements to those obtained for a quasipure BEC~\cite{Etrych:2025}, and find that the spectral shift in a thermal gas is suppressed compared to the BEC and, counterintuitively, near unitarity the spectra narrow with increasing $T$. Near $\Tc$, the spectra retain clear many-body character, with three key signatures:
(i)~the spectral shifts and widths are set by the energy scale $E_n=\hbar^2\knb^2/(4m_{\rm r})$, where $m_{\rm r}$ is the reduced mass of an impurity and a bath atom (in our equal-mass case $m_{\rm r}=m/2$, where $m$ is the atom mass), (ii)~the attractive branch extends across the resonance, and (iii)~the spectral width is asymmetric in $1/(\knb a)$, reaching a maximum value for $1/(\knb a)\approx0.4$. For higher $T/\Tc$, the system approaches the Boltzmann-gas regime: 
the attractive and repulsive spectral branches connect across the resonance, and the width is symmetric in $\lambda/a$, reaching its maximum (unitarity-limited) value at \mbox{$\lambda/a\approx0$}.
Throughout, we compare our results to bosonic functional determinant approach (FDA) calculations for the ideal Bose polaron\cite{Drescher:2021,Drescher:2024}~(see also \cite{Dzsotjan:2020,Schmidt:2018b}), finding good qualitative agreement.

Our experiments are performed using a gas of $^{39}$K atoms in an optical box trap~\cite{Gaunt:2013,Eigen:2016,Navon:2021} of volume $V= 3.3(7)\times 10^4\,\upmu$m$^3$, initially spin-polarized in the $\ket{F,m_F}=\ket{1,-1}$ state, here denoted as $\ket{\uparrow}$.
We prepare thermal samples with $T/\Tc$ up to $\approx 4$, varying both the temperature and the density of the sample\footnote{To avoid trapping atoms outside of the optical box, we always start our measurements with a quasipure BEC in a shallow trap of depth $\UD\approx\kB \times 50$\,nK. We then raise the depth to $\UD\approx\kB \times 1\,\upmu$K, inject energy by sinusoidally forcing the system using a magnetic field gradient, and allow the gas to rethermalize at $a_{\rm b}\approx100a_0$.}.
As illustrated in Fig.~\ref{fig1}(a) and following ~\cite{Etrych:2025}, we perform rf injection spectroscopy by coupling $\ket{\uparrow}$ to the $\ket{1,0}=\ket{\downarrow}$ (impurity) state with long, weak pulses~\footnote{
To minimize Fourier broadening and stay in the linear response regime, here we use pulse durations $\trf=\{0.8-6.4\}\,$ms and always limit the injection fractions to $\lesssim 10\%$.} of frequency $\omega/(2\pi)$, and measure the fraction of atoms transferred to $\ket{\downarrow}$ as a function of the detuning~$\delta/(2\pi)$ from the bare atomic transition frequency $\omega_0/(2\pi)$, such that $\omega=\omega_0+\delta$. We tune the impurity-bath interactions by changing the magnetic field in the proximity of the interstate Feshbach resonance at $\Bres=526.16(3)\,$G~\cite{Etrych:2025}; the intrabath interactions are always weakly repulsive, with the intrastate $(\ket{\uparrow}-\ket{\uparrow})$ scattering length $\ab\approx20a_0$.
To measure the injected impurity population, at the end of each experimental sequence we quench
$B\to \Bres$ and wait for all impurities to be lost, recording the total loss fraction $\Delta N/N$.

The theoretical spectra are calculated within the FDA for the ideal Bose polaron, a stationary (infinite-mass) impurity in a noninteracting Bose gas~(see Ref.~\cite{Drescher:2024} and \nameref{app:A} for further details on the calculations).
The calculations are exact for this limit and include an arbitrary number of excitations of the bath (capturing macroscopic deformations of it), but the model itself applies only approximately to our experiment; in particular, it neglects short-range bath correlations due to intrabath repulsion and/or three-body physics, and it naturally cannot capture impurity motion (which also induces bath correlations). 
To compare to this theory, we set $m_{\rm r}$ to our value $m/2$. 

\begin{figure*}[t!]
\centerline{\includegraphics[width=1.\textwidth]{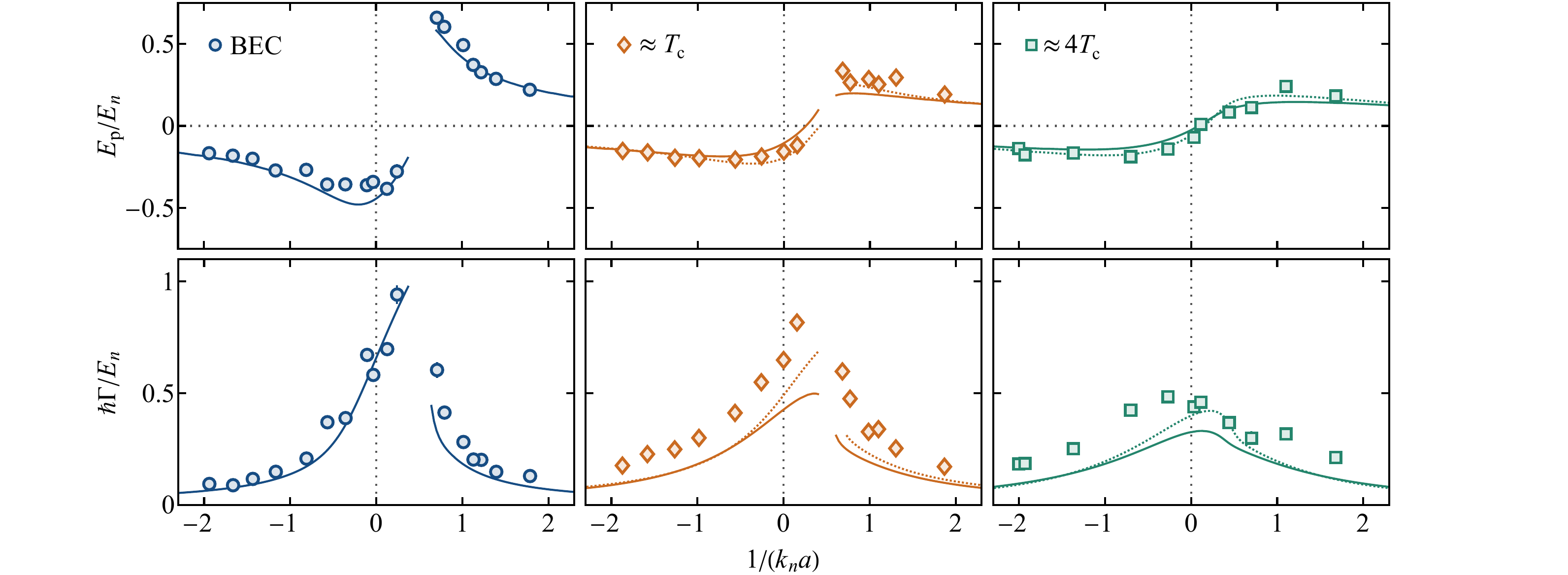}}
\caption{Spectrum mode $\Ep/\Enb$ (top) and half-width-at-half-maximum $\hbar\Gamma/\Enb$ (bottom) versus $1/(\knb a)$ for our three different baths (increasing $T/\Tc$ left to right); the solid lines show the corresponding ideal-polaron FDA calculations.
The dotted lines show ideal-polaron FDA calculations at half the bath temperature, so that $T/E_n$ is the same as in the experiment.
We refrain from extracting $\Ep$ and $\Gamma$ for the attractive branch for $1/(k_na)\gtrsim0.3$.
}
\label{fig2}
\vspace{-1em}
\end{figure*}

In Fig.~\ref{fig1}(b), we show an overview of our main results, the injection spectra $I(\delta)$ in both experiment (top) and theory (bottom), for various $1/(\knb a)$ and three different baths:
an initially quasipure BEC ($T/\Tc<0.1$), a Bose gas near criticality (\mbox{$T/\Tc\approx1$}), and a thermal gas (\mbox{$T/\Tc\approx4$})~\footnote{Note that in the ideal-polaron theory, the spectra are expected to evolve continuously as the bath temperature is increased, with no sharp change at $\Tc$, and we have experimentally checked that the injection spectra do not qualitatively change just below $\Tc$.}; the spectra are normalized such that $\int I(\delta){\rm d}\delta=1$, and for the BEC comparison, we use $T/\Tc= 0.15$ in the theory~(see \nameref{app:B}). For the BEC case, the attractive branch extends to $1/(\knb a)>0$ and for $1/(\knb a)>0.5$, both the attractive ($\delta<0$) and the repulsive ($\delta>0$) branches are clearly resolved.
For $T/\Tc\approx1$, the spectral shifts are generally suppressed compared to the BEC case. Across the resonance, spectral weight is more continuously transferred between the two branches, but the attractive branch still extends to $1/(\knb a)>0.5$. For $T/\Tc\approx4$, the spectral shift near unitarity is fully suppressed, the two branches connect across the resonance, and there is little attractive-branch spectral weight for $1/(\knb a)>0$.

The ideal-polaron calculations reproduce the qualitative features of the experimental spectral functions, with the exception of the attractive branch at $1/(\knb a)>0.5$, where the theory predicts multiple well-separated peaks, while a single broad feature is observed in the experiment.
In the theory, the multiple peaks correspond to the impurity binding with multiple bath atoms, with binding energies that are integer multiples of the single-boson binding energy (see also~\cite{Shchadilova:2016}).
However, this behavior is known to be pathological; for mobile impurities and realistic intrabath interactions, the short-range bath correlations are expected to break the simple integer scaling and suppress the weight of the lower-lying states~\cite{Levinsen:2021}.
The inclusion of these effects would likely improve the agreement with the experiment, but a quantitative understanding of these states remains a challenge~\cite{Etrych:2025}~(see also \cite{Chuang:2025,Robens:2026}).

In Fig.~\ref{fig1}(c), we show example spectra for our three $T/\Tc$ at different $1/(\knb a)$.
The temperature effects are the most striking in the range \mbox{$0\lesssim 1/(k_n a)\lesssim0.7$}, where the BEC spectra are the broadest.
For $T/\Tc\approx1$, the widths of the spectra are comparable to or smaller than for the BEC, and for $T/\Tc\approx4$ the spectra are significantly more narrow. This behavior is notably different than the temperature evolution of spectral features in cold Fermi gases, where broadening with increasing temperature is generally observed, and only at temperatures above the Fermi temperature the spectra slowly narrow as $1/\sqrt{T}$ due to unitarity-limited collisions~\cite{Yan:2019,Mukherjee:2019}.

We characterize the spectra by their peak position $\Ep/\Enb$ and half-width-at-half-maximum $\hbar\Gamma/\Enb$~\footnote{We always correct $\Gamma$ for residual Fourier broadening arising from our finite rf pulse times, $\trf$, using \mbox{$\Gamma=\left(\Gamma_{\rm e}^2-\Gamma_{\rm t}^2\right)^{1/2}$, where $\Gamma_{\rm e}$} is the raw extracted width and \mbox{$\Gamma_{\rm t}\approx2.78/\trf$} is the width of our response function\cite{Etrych:2025}.}, and in Fig.~\ref{fig2}, we plot these extracted quantities as a function of $1/(\knb a)$.
For $T/\Tc\approx 1$, the measured $|\Ep|/E_n$ values are significantly lower than for the BEC, but they change only slowly between $T/\Tc\approx1$ and $4$.
The width at $T/\Tc\approx1$ shows the same asymmetric dependence on $1/(\knb a)$ as the BEC data [with a maximum at $1/(\knb a)>0$], but it becomes more symmetric for $T/\Tc\approx4$, with the maximum at $1/(\knb a)\approx0$.
The asymmetry between the attractive and repulsive branches at low $T$ is a many-body effect, due to bound states that exist only for $a>0$. In the high-$T$ limit, only the two-body scattering states are relevant and the situation for positive and negative $a$ is symmetric.

While the BEC data are well captured by the ideal-polaron theory (solid lines, see also \nameref{app:B}), the agreement is only qualitative for the thermal baths; in particular, the experimental widths are about $50\%$ larger than the theoretical ones, and for $T/\Tc\approx1$, the experimental $\Ep/E_n$ of the repulsive branch is about $30\%$ larger than the theoretical one. 

These differences could arise for at least three reasons: 
(i) the motion of the impurity induces correlations between bath atoms, which could broaden the spectra,~ (ii) the ratio of thermal and interaction energy scales is intrinsically different in the ideal-polaron model (where $\kB \Tc \approx0.88E_n$) and our mobile-impurity case (where $\kB \Tc\approx0.44E_n$), leading to more significant suppression of (many-body) interaction effects in the theory compared to the experiment; for comparison, in Fig.~\ref{fig2} we also show ideal-polaron FDA calculations at half the bath temperature (dotted lines), so that $T/E_n$ is the same as in the experiment,~(iii) for rf injection starting from a spin-polarized thermal Bose gas, one expects the spectral shifts to be enhanced due to bosonic bunching, which arises from coherence between the impurity and bath states. In the mean-field limit (at low $|a|$), one expects $E_{\rm mf}=2\pi \hbar^2 \alpha n a/m_{\rm r}$ with $\alpha=2$ (compared to $\alpha=1$ for a pure BEC)~\cite{Harber:2002,Zwierlein:2003b}, and such enhancement could still play a role in the strongly interacting regime. 
However, note that even for low $a$, we curiously find $\alpha$ closer to $1$~(see~\nameref{app:C}), and understanding the role of coherence in injection experiments remains a challenge for the future.

\begin{figure}[t!]
\centerline{\includegraphics[width=1\columnwidth]{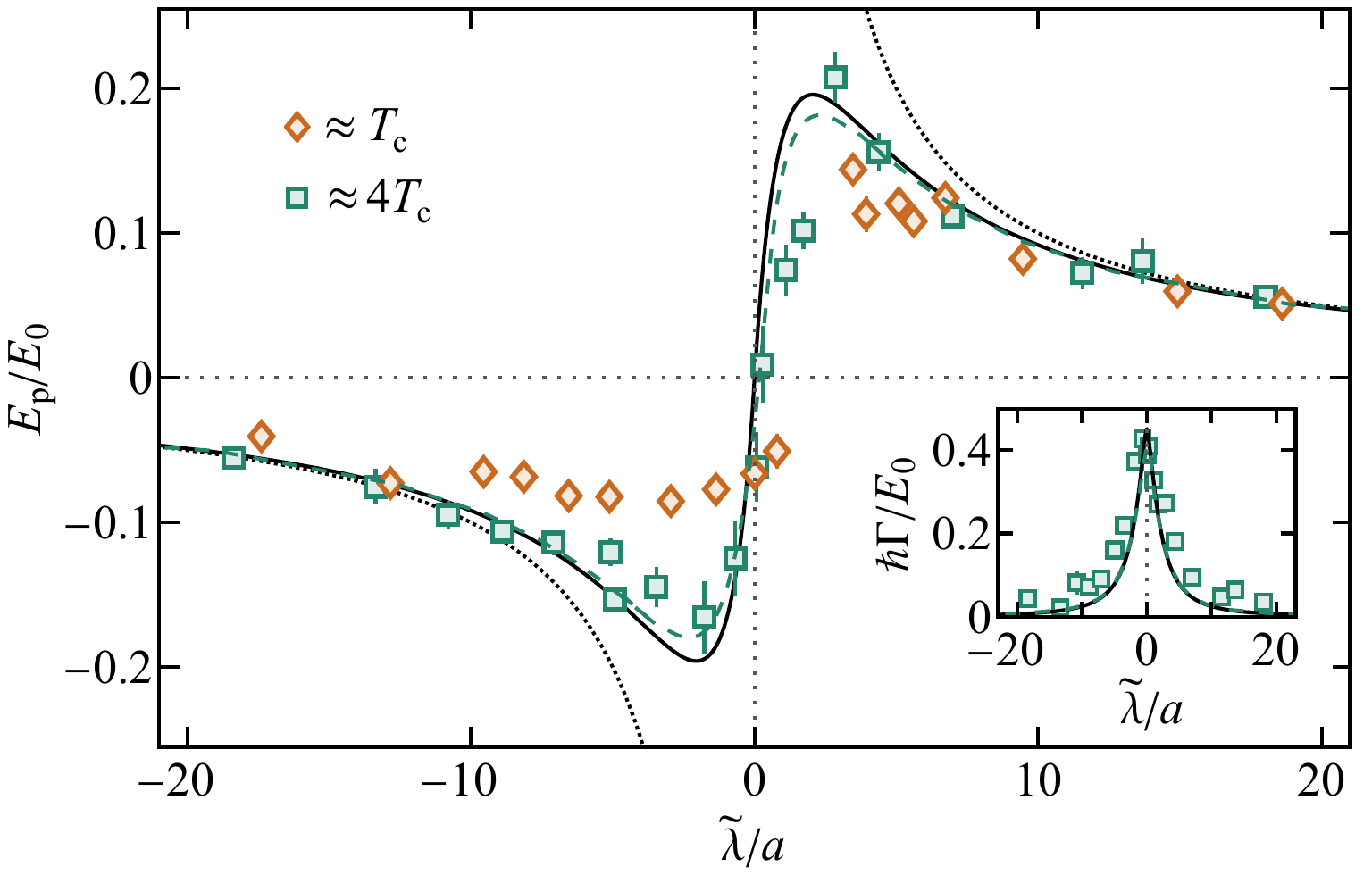}}
\caption{
Crossover to the Boltzmann-gas regime.
The main panel shows $\Ep/E_0$ versus $\tilde{\lambda}/a$, where $E_0 =2\pi \hbar^2\tilde{\lambda} \alpha n/m_{\rm r}$ and $\tilde{\lambda}=\lambda \sqrt{m/(2 m_{\rm r})}$.
The inset shows $\hbar\Gamma/E_0$ versus $\tilde{\lambda}/a$ for $T/\Tc\approx4$.
The solid lines show the universal two-body prediction valid for $T/\Tc\gg1$, while the dotted line shows the mean-field prediction $\Ep=E_{\rm mf}$.
For reference, the dashed lines show the ideal-polaron theory for $T/T_{\rm c}=4$, which for $T \to \infty$ coincides with the universal two-body theory.
}
\label{fig3}
\vspace{-1em}
\end{figure}

Finally, we focus on the crossover to the classical Boltzmann-gas regime, in which only two-body effects are relevant and one can (to a good approximation) obtain the spectral shift and width, respectively, from the real and imaginary parts of the complex energy 
$\Sigma=\Ep-i\hbar\Gamma=-2\pi\hbar^2 \alpha n f(a)/m_{\rm r}$, where $f(a)=\braket{-a/(1+ika)}_{\rm th}$ is the thermally averaged two-body scattering amplitude~(see~\nameref{app:D}).
In this case, the natural interaction parameter is $a/\tilde{\lambda}$, where the mass-adjusted $\tilde{\lambda}=\lambda\sqrt{m/(2m_{\rm r})}$ and $\lambda=\sqrt{2\pi\hbar^2/(m\kB T)}$, and the natural energy scale is $E_0=E_{\rm mf}\, \tilde{\lambda}/a=2\pi\hbar^2\tilde{\lambda}\alpha n /m_{\rm r}$.
Near the resonance, the interactions become unitarity-limited: at $\lambda/a=0$, the scattering amplitude and $\Sigma$ become purely imaginary, the spectral shift is zero, and the width is maximal.

In Fig.~\ref{fig3}, we compare our measurements to this universal high-$T$ prediction (solid line), plotting $\Ep/E_0$ versus $\tilde\lambda/a$ for our thermal baths; we experimentally determine $E_0$ from the slope of $\Ep$ versus $a$ in the mean-field regime ($|a|\ll \lambda$)~(see \nameref{app:C}). The $\Ep$ data approach the theory for our highest temperature $T/\Tc\approx4$; here the width $\hbar\Gamma/E_0$ also agrees with the theory for $\lambda/a\approx0$, but is systematically higher away from unitarity (see inset).
This temperature corresponds to a phase-space density $n\lambda^3\approx0.3$, suggesting that the two-body theory becomes valid only for an essentially nondegenerate bath, while for a degenerate bath near $\Tc$ (where all three principal lengthscales $a$, $\lambda$, and $1/\knb$ are comparable in magnitude) many-body effects are significant~\footnote{Note that similar conclusions were obtained in interferometric experiments with single-component Bose gases~$\cite{Fletcher:2017}$; for $n\lambda^3=0.13$ ($T/\Tc\approx7$) the measurements were consistent with the two-body theory, but measurements for $n\lambda^3=0.54$ ($T/\Tc\approx3$) revealed effects of three-body correlations.}.
For reference, we show the ideal-polaron theory at $T/T_{\rm c}=4$ (dashed lines), which is also already close to the two-body theory (see also \nameref{app:D}).

In conclusion, we have investigated the spectral response of impurities strongly interacting with a thermal bosonic bath, revealing the importance of many-body effects in the degenerate regime, and studying the crossover to the classical Boltzmann-gas regime with unitarity-limited scattering.
The comparison of our thermal-bath spectra with previous measurements for an impurity in a quasipure BEC highlights the conceptual differences between bosonic and fermionic baths: while sharp quasiparticle features emerge at low $T$ in the fermionic case, the bosonic spectral response is intrinsically broad at strong interactions.
The qualitative observations are well captured in the ideal-polaron theory, which neglects impurity motional degrees of freedom and short-range bath correlations, but crucially includes an arbitrary number of bath excitations and can thus capture the macroscopic deformation of the bath.

In the future, it would be interesting to extend our study to the case of light impurities, where Efimov physics is predicted to play a more important role~\cite{Sun:2017,Christianen:2024}, or to systems with stronger intrabath repulsion, where the quasiparticles are expected to be more well-defined~\cite{Guenther:2021} (see also~\cite{Henke:2025,Alhyder:2026}) and, furthermore, the thermal bath becomes hydrodynamic~\cite{Hilker:2022}.
It would also be interesting to vary both $n$ and $T$ at different fixed values of $T/\Tc$ and to explore impurity concentration effects~\cite{Baroni:2024}.
Our work motivates the development of novel theoretical approaches that capture the effect of short-range bath correlations [especially relevant for the attractive branch at $1/(\knb a)>0$] at arbitrary $T$.

We thank Martin Gazo, Richard Schmidt, and Meera Parish for discussions and comments on the manuscript.
The Cambridge work was supported by EPSRC [Grant Nos.~EP/P009565/1 and EP/Y01510X/1], ERC [UniFlat], and STFC [Grants No.~ST/T006056/1 and No.~ST/Y004469/1]. Z.H. acknowledges support from the Royal Society Wolfson Fellowship.
The Heidelberg work was supported by the DFG (German Research Foundation) under Project No.~273811115 (SFB
1225 ISOQUANT) and under Germany’s Excellence Strategy EXC2181/1-390900948 (the Heidelberg STRUCTURES Excellence Cluster).
The data that support the findings of this article are openly available~\footnote{\href{https://doi.org/10.17863/CAM.134561}{https://doi.org/10.17863/CAM.134561}.}.

\renewcommand\theequation{A\arabic{equation}}

\appendix

% % \vspace*{-0.5em}
% \begin{center}
%    {\large\bfseries End Matter}
% \end{center}

\newpage
%%%%%%%%%%%%%%%%%%%%%%%%%%%%%%%%
%%%%%%%%%%%%%%%%%%%%%%%%%%%%%%%%

\onecolumngrid
\section{\textsc{End Matter}}
%\vspace{-1em}
\twocolumngrid
\subsection[Appendix A]{Appendix A: Functional determinant approach}
\label{app:A}
%\vspace{-0.5em}
Following Refs.~\cite{Drescher:2020,Drescher:2021,Drescher:2024} (see also~\cite{Schmidt:2018b,Dzsotjan:2020}), for a mobile equal-mass impurity in a Bose gas, one can
write the Hamiltonian in the frame comoving with the impurity as
\begin{equation}
  \hat H = \sum_i \frac{\hat{\mathbf p}_i^2}{2m_{\rm r}} + \sum_i V_{\rm IB}(\hat{\mathbf x}_i)
  + \frac{:(\sum_i\hat{\mathbf p}_i - \mathbf p_0)^2:}{2\,m}\,,
\end{equation}
where $\mathbf p_0$ is the impurity momentum, $\hat{\mathbf p}_i$ and $\hat{\mathbf x}_i$ denote the momentum and position operators of the $i$-th bath boson, and $V_{\rm IB}$ is the impurity-boson potential centered around $\mathbf x=0$ in the comoving frame.
The third term on the r.h.s.~accounts for impurity recoil and generates effective higher-order correlations between bath particles; $:\ldots:$ denotes normal ordering.
Throughout we set this third term to zero; note that this approximation is exact in the high-temperature, two-body and weak-interactions limits.
The real-time response of the system is then given by
\begin{equation}
  S(t) = \langle e^{i\hat H_0t/\hbar} e^{-i\hat Ht/\hbar} \rangle\,,
\end{equation}
where $\hat H_0$ denotes the Hamiltonian with $V_{\rm IB}=0$.
This response has contributions from the condensate that is present for $T < T_{\rm c}$, from the thermal modes for $T > 0$, and from the correlations between condensate and thermal modes for $0 < T < T_{\rm c}$; see~\cite{Drescher:2024} for the explicit expressions.
Finally, the measured rf spectrum is obtained as the Fourier transform
\begin{equation}
  I(\delta) = \frac1\pi {\rm Re} \int_0^\infty \, S(t) e^{i\delta t}  {\rm d}t.
\end{equation}

\vspace{1em}
\subsection[Appendix B]{Appendix B: Effect of nonzero $T$ for impurities in a BEC}
\label{app:B}
\begin{figure}[b!]
\centerline{\includegraphics[width=1.0\columnwidth]{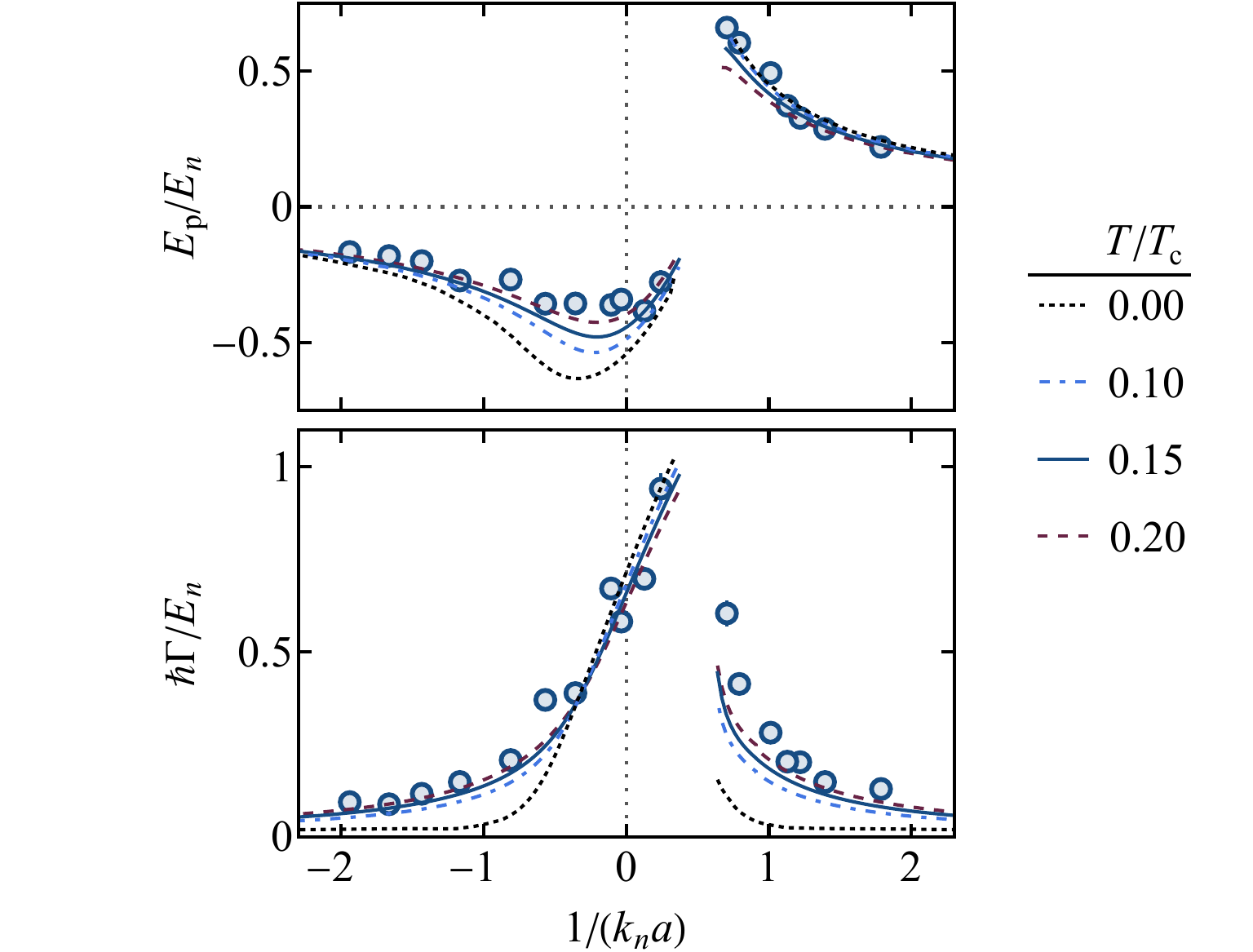}}
\caption{
Comparison of our BEC-bath data~\cite{Etrych:2025} with the ideal-polaron theory at zero and nonzero $T$. Plot of $\Ep/E_n$ (top) and $\hbar\Gamma/E_n$ (bottom) versus $1/(k_n a)$, showing the data (symbols) and the ideal-polaron theory at different $T/\Tc$ (legend). 
}
\label{fig4}
\end{figure}

In Fig.~\ref{fig4}, we illustrate the effect of a small nonzero $T$ in the theory (see also~\cite{Drescher:2024}). We plot the ideal-polaron theory for $T=0$ and for different small $T/\Tc$, alongside our BEC data. The theoretical $\Ep$ values for the repulsive branch show only small variation with $T$, while $|\Ep|$ of the attractive branch for $-1\lesssim1/(k_na)\lesssim0$ is suppressed with increasing $T$. 
For $1/(k_n|a|)\gtrsim1$, the theoretical $\Gamma$ is very small only for $T=0$, and any small $T$ causes significant broadening.
On the other hand, near $1/(k_na)\approx0$, $\Gamma$ changes only slowly with $T$.
We find that $T/\Tc=0.15$ captures the data reasonably well for all $1/(k_n a)$.

Initially, our quasipure BEC has $T\approx 10\,$nK~\cite{Lopes:2017b} such that $T/\Tc<0.1$, but both the impurities and the bath can be excited during injection, due to dynamics arising from inhomogeneity near the box edges and due to three-body losses. The effect of such excitations should be captured by an elevated (effective) $T$, and for the theory comparisons in Figs.~\ref{fig1} and \ref{fig2} we used $T/\Tc=0.15$.
\vspace{1em}

\begin{figure}[b!]
\centerline{\includegraphics[width=1.0\columnwidth]{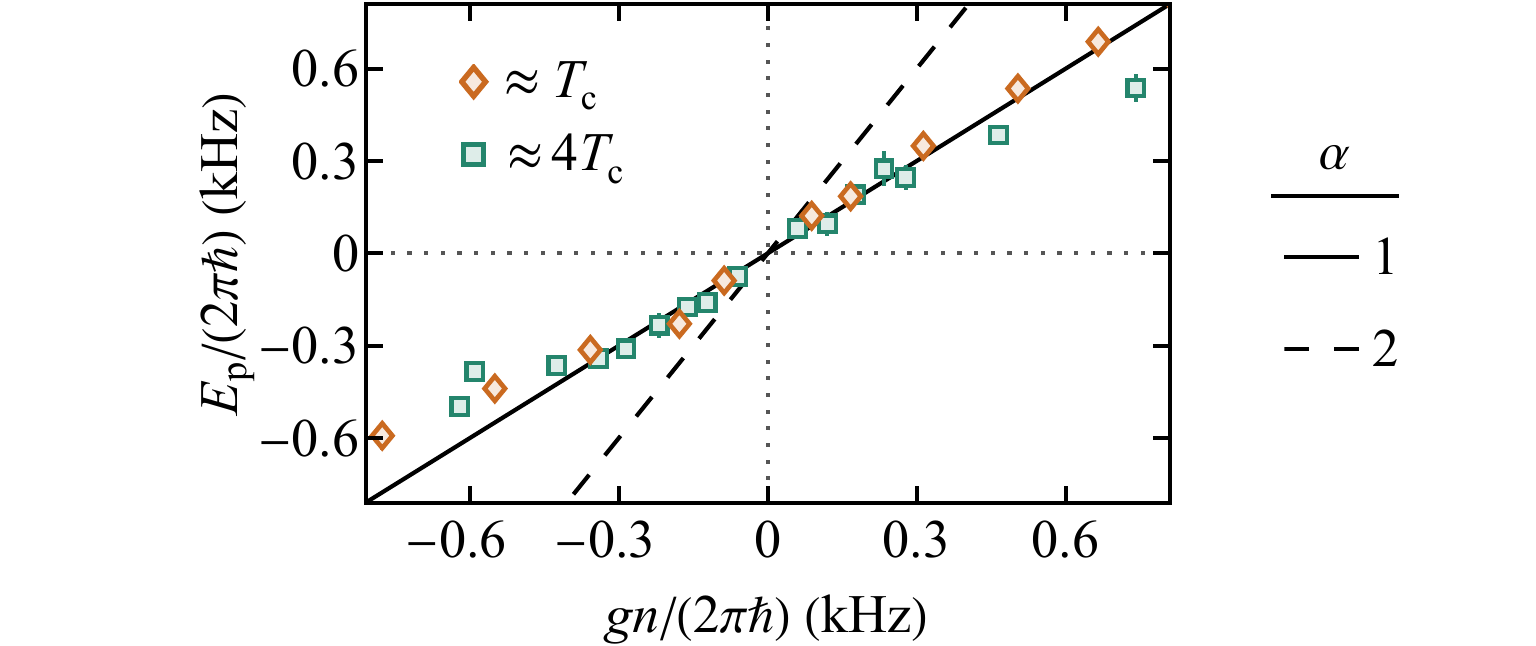}}
\caption{
Mean-field energy shift in the weakly interacting regime. Spectrum mode $\Ep$ versus $gn$, where $g=2\pi\hbar^2a /m_{\rm r}$. The solid and dashed lines show $\Ep=\alpha gn$ with, respectively, $\alpha=1$ and $2$ (see text for details).
}
\label{fig5}
\end{figure}

%\vspace{-0.5em}
\subsection[Appendix C]{Appendix C: Spectral shifts in the weakly interacting regime}
\label{app:C}
\vspace{-0.5em}
In Fig.~\ref{fig5}, we show our measured $\Ep$ values for low $|a|$, where spectral shifts reflect the mean-field impurity energy $E_{\rm mf}=2\pi \hbar^2 \alpha n a/m_{\rm r}$. For our BEC data~\cite{Etrych:2025}, we used such measurements to calibrate the bath density. Here, for our thermal baths, we directly obtain the bath atom number $N$ and temperature $T$ from time-of-flight absorption images, and calculate the density using the independently measured $V$~\cite{Gaunt:2013}.

For injection spectroscopy of a thermal gas, one expects $\alpha=2$, reflecting the enhanced probability of finding two identical bosons on top of each other in the initial single-component gas~\cite{Harber:2002,Zwierlein:2003b}; for a BEC, on the other hand, \mbox{$\alpha=1$}. 
However, fitting our data in Fig.~\ref{fig5} gives $\alpha=1.1(4)$, with the error dominated by our systematic uncertainties in $N$ and $V$.
Experimentally, the measured $\alpha$ should be slightly lower than $2$ due to the nonzero fraction of transferred impurities, but since our transfer fractions are $\lesssim10\%$, this explains only a small part of the discrepancy.

\renewcommand\theequation{D\arabic{equation}} 
%\vspace{-0.5em}
\subsection[Appendix D]{Appendix D: Two-body theory valid for $T/T_{\rm c}\gg1$}
\label{app:D}
%\vspace{-0.5em}
In the high-$T$ limit, for any impurity mass, the ideal-polaron theory coincides with the universal two-body theory, which gives the spectral shift and width, respectively, from the real and imaginary parts of the complex energy $\Sigma=\Ep-i\hbar\Gamma=-2\pi\hbar^2 \alpha n f(a)/m_{\rm r}$, where
\begin{equation}
    f(a) = -\frac{\tilde{\lambda}^3}{2^{3/2}\pi^3}\int \frac{a}{1+ika} \exp\left(-\frac{\hbar^2k^2}{2m_{\rm r} \kB T}\right){\rm d}^3k
\end{equation}
is the thermally averaged scattering amplitude. More explicitly, one obtains the universal functions:
\begin{align}
  \frac{\Ep}{E_0} & = \frac{v}{\pi} \left[1 - \frac{|v|}{\sqrt{2}} \exp\left(\frac{v^2}{2\pi}\right)\, \text{erfc}\left(\frac{|v|}{\sqrt{2\pi}}\right)\right]\,,
  \\
  \frac{\hbar\Gamma}{E_0} & = \frac{\sqrt{2}}{\pi}\left[1+ \frac{v^2}{2\pi} \exp\left(\frac{v^2}{2\pi}\right)\, \text{Ei}\left(\frac{-v^2}{2\pi}\right)\right]\,,
 \label{eq:Gamma} 
\end{align}
where $v=\tilde{\lambda}/a$ and $\text{Ei}(x)$ denotes the exponential integral.

\end{document}